\documentclass{ws-procs9x6}
\newcommand{\dif}{\text{d}}
\newcommand{\mypar}[1]{\frac{\partial}{\partial #1}}

\DeclareMathOperator{\Tr}{Tr}
\DeclareMathOperator{\tr}{tr}

\renewcommand{\tilde}[1]{\widetilde{#1}}
\let\Re\undefined
\DeclareMathOperator{\Re}{Re}
\let\Im\undefined
\DeclareMathOperator{\Im}{Im}

\begin{document}

\title{Scalar Casimir Energies for Separable Coordinate Systems:
Application to Semi-transparent Planes in an Annulus}

\author{J.~Wagner$^*$ and K.~A.~Milton$^\dagger$}

\address{University of Oklahoma,\\
Homer L. Dodge Department of Physics and Astronomy,\\
Norman, OK, 73019.\\
$^*$E-mail: wagner@nhn.ou.edu\\
$^\dagger$E-mail: milton@nhn.ou.edu}

\author{K.~Kirsten$^\ddagger$}

\address{Baylor University\\
Department of Mathematics\\
One Bear Place \# 97328\\
Waco, TX 76798-7328\\
$^\ddagger$E-mail: Klaus\_Kirsten@baylor.edu}

\begin{abstract}
We derive a simplified general expression for the two-body scalar
Casimir energy in generalized separable coordinate systems. We apply this
technique to the case of radial semi-transparent planes in the annular region
between two concentric Dirichlet cylinders. This situation is explored both
analytically and numerically.
\end{abstract}


\bodymatter

\section{Introduction}
In 1948 Casimir\cite{Casimir:1948dh}  predicted that two parallel
perfectly reflecting mirrors would attract each other with a pressure
of $P= \pi^2/240 a^4$. Since then much work has been done studying a
variety of geometries and materials. Much of this work has been
summarized and referenced in review articles by M.~Bordag {\it et al}
\cite{Bordag:2001qi} and K.~A.~Milton\cite{Milton:2004ya}, and more
completely in two books by the same authors
\cite{Milton:2001yy,Bordag:2009a}.

This work only concerns itself with the Casimir effect for a massless
scalar field. In order to proceed we will start with
the multiple scattering expression for the Casimir energy
\begin{equation}\label{intro:TrlnGVGV}
  E=\frac{1}{4\pi}\int\limits_{-\infty}^\infty\dif\zeta \Tr
  \ln(1-G_1V_1G_2V_2).
\end{equation}
Here $\zeta$ is the imaginary frequency, and $G_i$ is the Green's function
referring to a single potential $V_i$.
An equivalent expression was first used by Renne\cite{Renne:1971a} in 1971, and
more recently by many others \cite{Emig:2002xz, Bulgac:2005ku,
Emig:2006uh}. A very good derivation is given by Kenneth and
Klich\cite{Kenneth:2007jk}.

\section{Separation of Variables}
Equation \eqref{intro:TrlnGVGV} is a fairly complicated formula to work with. 
We have to perform a 3-dimensional trace
of the logarithm of the $1-G_1V_1G_2V_2$ operator. We also have to
solve a partial differential equation to find $G_1$ and
$G_2$. However, by working in a coordinate system in which the
Helmholtz equation is separable we can greatly simplify this approach.
The result will allow us to move the trace inside the logarithm, where it
will become a simple integral, and we will only have to solve an ordinary
differential equation to find a reduced Green's function for a single
coordinate.

In this section we will find a simplified expression based on a
general separation of variables using the St\"{a}ckel determinant. We
will follow the notation of Morse and Feshbach\cite{Morse:1953a}.

We write the Green's function as a sum of eigenfunctions times a
reduced Green's function,
\begin{equation}\label{sep:G}
  G(\vec{x},\vec{x}')=\sum_{\alpha_2}\sum_{\alpha_3}
  \frac{\rho}{M_1f_2f_3}
  \chi_2(\xi_2)\chi_3(\xi_3)
  \chi_2(\xi'_2)\chi_3(\xi'_3) g(\xi_1,\xi_1').
\end{equation}
The $M_1(\xi_2,\xi_3)$ is the minor of the St\"{a}ckel determinant,
and the $f_i(\xi_i)$ functions are functions of a single variable
related to the scale factors of the generalized coordinate system as
defined in Morse and Feshbach.\cite{Morse:1953a}  The $\chi_2(\xi_2)$
and $\chi_3(\xi_3)$ and $\alpha_2$ and $\alpha_3$ are the
eigenfunctions and eigenvalues determined by the simultaneous set of
equations,
\begin{subequations}\label{sep:efe}
\begin{eqnarray}
  \left(-\frac{1}{f_2}\mypar{\xi_2}f_2\mypar{\xi_2}+
  \Phi_{21}\zeta^2+\Phi_{22}\alpha_2^2+\Phi_{23}\alpha_3^2\right)
  \chi_2(\xi_2;\zeta,\alpha_2,\alpha_3)=0,\\
  \left(-\frac{1}{f_3}\mypar{\xi_3}f_3\mypar{\xi_3}+
  \Phi_{31}\zeta^2+\Phi_{32}\alpha_2^2+\Phi_{33}\alpha_3^2\right)
  \chi_3(\xi_3;\zeta,\alpha_2,\alpha_3)=0.
\end{eqnarray}
\end{subequations}
The $\chi$ eigenfunctions are orthogonal with respect to some weighting
function $\rho(\xi_2,\xi_3)$,
\begin{equation}\label{sep:orth}
  \int \dif \xi_2 \dif \xi_3 \rho 
  \chi_2(\alpha_2,\alpha_3)\chi_2(\alpha_2',\alpha_3')
  \chi_3(\alpha_2,\alpha_3)\chi_3(\alpha_2',\alpha_3')
  =\delta_{\alpha_2,\alpha_2'} \delta_{\alpha_3,\alpha_3'}.
\end{equation}
Using \eqref{sep:efe}, we find that the reduced
Green's function in \eqref{sep:G} satisfies the
differential equation in the single remaining coordinate,
\begin{multline}\label{sep:red_g}
  \bigg(-\frac{1}{f_1}\mypar{\xi_1}f_1\mypar{\xi_1}+
  \Phi_{11}\zeta^2\\+\Phi_{12}\alpha_2^2+\Phi_{13}\alpha_3^2
  +v(\xi_1)\bigg)g(\xi_1,\xi_1';\zeta,\alpha_2,\alpha_3)=
  \frac{\delta(\xi_1-\xi_1')}{f_1}.
\end{multline}

Working with the Casimir energy written as \eqref{intro:TrlnGVGV}, by
expanding the log we can write
\begin{equation}
  E=-\frac{1}{4\pi}\int\limits_{-\infty}^\infty \dif \zeta
  \sum_{s=1}^\infty \frac{1}{s}\Tr(G_1V_1G_2V_2)^s.
\end{equation}
The simplification comes if the potentials are  functions of only the
single coordinate $\xi_1$, with the form
$V_i(\vec{x})=v_i(\xi_1)/h_1^2$. The scale factor $h_1$ is exactly what
is needed to apply the orthogonally condition \eqref{sep:orth} in
performing the trace. Finally if the potential consists of two separate
non-overlapping potentials, we can show
\begin{equation}
  \Tr(G_1V_1G_2V_2)^s=\sum_{\alpha_2,\alpha_3} \tr(g_1v_1g_2v_2)^s
  =\sum_{\alpha_2,\alpha_3} \left(\tr g_1v_1g_2v_2\right)^s.
\end{equation}
 The interaction Casimir energy can now be written in general
 separable coordinates as
\begin{equation}\label{sep:en}
  E=\frac{1}{4\pi}\int\limits_{-\infty}^\infty\dif\zeta
  \sum_{\alpha_2,\alpha_3}\ln(1-\tr g_1v_1g_2v_2).
\end{equation}

\section{Casimir Energy for Planes in an Annular Cavity}
\begin{figure}
\begin{minipage}[t]{0.47\linewidth}
  \begin{center}
    \includegraphics{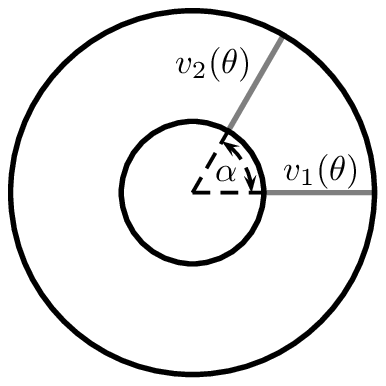}
    \caption{\label{fig:ann} An annulus with inner radius $a$, outer
      radius $b$, and two semitransparent potentials at $\theta=0$ and
      $\theta=\alpha$.}
  \end{center}
\end{minipage}
\hfill
\begin{minipage}[t]{0.47\linewidth}
  \begin{center}
    \includegraphics{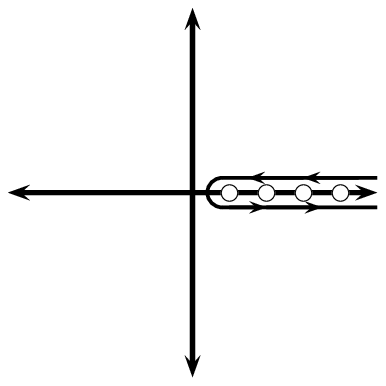}
    \caption{\label{fig:contour} The contour $\gamma$ is defined
      around the positive real line, while not enclosing zero.}
  \end{center}
\end{minipage}
\end{figure}

As an application we will proceed for the case of two semitransparent
radial planes in the region between two concentric cylinders, as shown in
figure \ref{fig:ann}.

This geometry is similar to the wedge geometry first studied in
1978\cite{Dowker:1978md,Deutsch:1978sc}, with a good review by Razmi
and Modarresi\cite{Razmi:2005qf}. However here we include circular
boundaries in addition to the wedge boundaries. We will enforce
Dirichlet boundary condition on the inner and outer cylinder. This is
similar to situations studied by Nesterenko {\it et al}
\cite{Nesterenko:2002jp, Nesterenko:2002ng} for global Casimir energies
for the case of one circular boundary and by Saharian {\it et al}
\cite{Saharian:2005dv, Saharian:2007sc} for the local
properties of the stress energy tensor for the case of both one and
two circular boundaries. The radial potentials will be semi-transparent
delta-function potentials in the angular coordinates,
$v_1(\theta)=\lambda_1\delta(\theta)$ and
$v_2(\theta)=\lambda_2\delta(\theta-\alpha)$. This is most similar to
the recent work by Brevik {\it et al}\cite{Brevik:2009vf,
Ellingsen:2009ff}, and Milton {\it et al}\cite{Milton:2009a}.

This problem can be solved using separation of variables, leaving 
$\xi_1$ as the azimuthal coordinate $\theta$. This means we will write
our reduced Green's function in the azimuthal coordinate, which is
different from the traditional way of writing the reduced Green's
function in terms of the radial coordinate. From equation \eqref{sep:en}
we can immediately write
\begin{equation}\label{ann:en_sum}
  \frac{E}{L_z}=\frac{1}{4\pi}\int\limits_0^\infty \dif\zeta\sum_\eta
  \ln(1-\tr g_\eta^{(1)}v_1g_\eta^{(2)}v_2).
\end{equation}
The Green's function is written in terms of exponential functions that,
due to the periodicity requirement, give the expression
\begin{equation}\label{ann:gvgv}
  \tr g_\eta^{(1)}v_1g_\eta^{(2)}v_2=
  \frac{\lambda_1 \lambda_2 \cosh^2\big(\eta(\pi-\alpha)\big)}
  {(2\eta \sinh \eta \pi + \lambda_1 \cosh \eta \pi)
    (2\eta \sinh \eta \pi + \lambda_2 \cosh \eta \pi)}.
\end{equation}

The $\eta$s are the eigenvalues of the modified Bessel
equation of purely imaginary order,
\begin{equation}\label{ann:eve}
  \left[-r \mypar{r}r\mypar{r}+\kappa^2 r^2\right]R_\eta(\kappa r)
  =\eta^2 R_\eta(\kappa r).
\end{equation}

Using the argument principle we can take a complicated sum over
eigenvalues and turn it into a contour integral around the real line as
shown in figure \ref{fig:contour}. For this we need a secular function
$D(\eta)$, which is analytic along the real line and has the value
zero at the eigenvalues. In this case we define $R_\eta(\kappa a)=0$
then the eigenvalue condition is given by $D(\eta)=R_\eta(\kappa
b)$. The eigenfunction $R_\eta$ can be written in terms of modified
Bessel functions
\begin{equation}\label{ann:R}
  R_\eta(\kappa r)=K_{i\eta}(\kappa a)\tilde{I}_{i\eta}(\kappa r)-
  \tilde{I}_{i\eta}(\kappa a)K_{i\eta}(\kappa r),
\end{equation}
where we define $\tilde{I}_{\eta}(x)$ as the part of the modified
Bessel function $I_\eta(x)$ even in $\eta$.

\begin{figure}
\begin{center}
\includegraphics{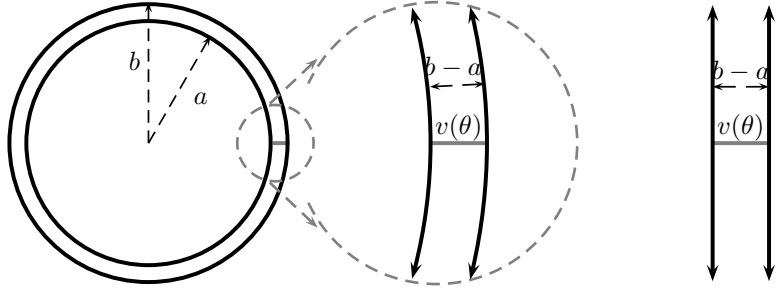}
\caption{\label{fig:limit} If the inner and outer radii are both large
in comparison to their separation, we should recover the case of a
rectangular piston.}
\end{center}
\end{figure}

The energy per unit length $L_z$ can be written as
\begin{multline}\label{ann:en_int}
  \frac{E}{L_z}=\frac{1}{8\pi^2i}\int\limits_0^\infty \kappa \dif \kappa
  \int_\gamma \dif \eta \left[\mypar{\eta} 
  \ln \big(K_{i\eta}(\kappa a)\tilde{I}_{i\eta}(\kappa b)-
  \tilde{I}_{i\eta}(\kappa a)K_{i\eta}(\kappa b)\big) \right]\\ \times
  \ln\left(1-
 \frac{\lambda_1 \lambda_2 \cosh^2\big(\eta(\pi-\alpha)\big)}
  {(2\eta \sinh \eta \pi + \lambda_1 \cosh \eta \pi)
    (2\eta \sinh \eta \pi + \lambda_2 \cosh \eta \pi)}\right).
\end{multline}
A quick check of this answer is to look at the limit of large inner
and outer radius, as shown in figure \ref{fig:limit}. This should then
give the answer for a rectangular piston.
For this limit we need the uniform asymptotic expansions of
$K_{i\eta}$ and $\tilde{I}_{i\eta}$, which are worked out by Dunster
\cite{Dunster:1990a,Olver:1974a}. We should also redefine our
dimensionless variables in terms of the dimensionful quantities that
will appear in the rectangular piston case, $\tilde{\eta}=\eta/a$,
$\tilde{\lambda}=\lambda/a$, and $d=\alpha a$. In this asymptotic
region we recover the formula for a rectangular piston,
\begin{multline}\label{ann:en_pist}
  \frac{E}{L_z}=\frac{1}{8\pi^2i}\int\limits_0^\infty \kappa \dif \kappa
  \int_\gamma \dif \tilde{\eta} \left[\mypar{\tilde{\eta}} 
  \ln \frac{\sin\left(\sqrt{\tilde{\eta}^2-\kappa^2}(b-a)\right)}
      {\sqrt{\tilde{\eta}^2-\kappa^2}} \right]\\ \times
  \ln\left(1-
  \frac{\tilde{\lambda}_1 \tilde{\lambda}_2 e^{-2\tilde{\eta} d}}
  {(2\tilde{\eta}+\tilde{\lambda}_1)(2 \tilde{\eta}+\tilde{\lambda}_2)}
  \right).
\end{multline}
The contour integral over
$\tilde \eta$  simply ensures that $\eta^2=\kappa^2+(m\pi/(b-a))^2$.

\section{Numerical Results for Dirichlet Planes}
\begin{figure}
\begin{minipage}[t]{0.47\textwidth}
  \includegraphics[width=\textwidth]{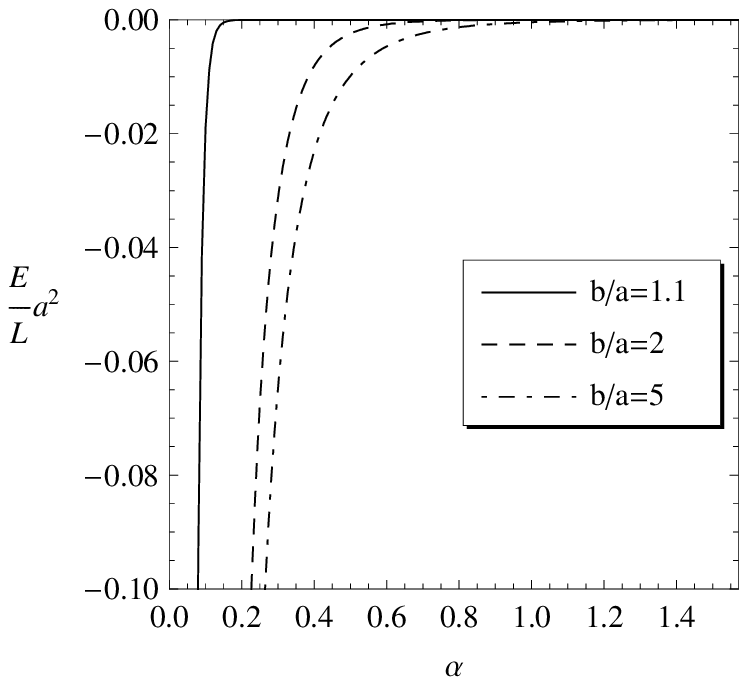}
  \caption{\label{ann:pi_4}This figure shows the energy per length vs
    the angle between the plates. The energy is scaled by the inner
    radius $a$.}
\end{minipage}
\hfill
\begin{minipage}[t]{0.47\textwidth}
  \includegraphics[width=\textwidth]{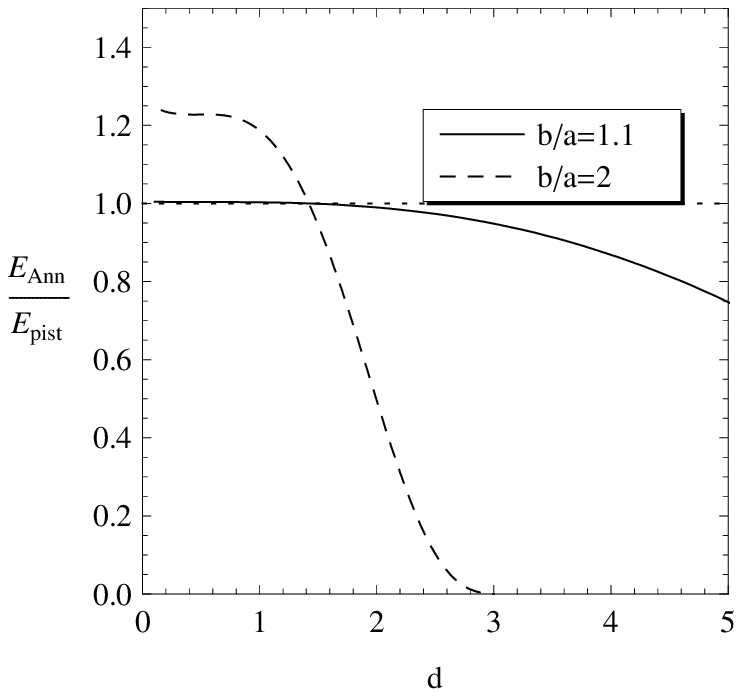}
  \caption{\label{ann:comparison}This figure shows the ratio of the
    energies of an annular piston to a rectangular piston of similar
    dimension vs average separation distance between the plates. The
    separation distance is scaled by the finite size of the piston
    $b-a$. For $b/a=2$ only the result for $\alpha\in [0,\pi]$ is shown.}
\end{minipage}
\end{figure}

The Casimir energy in equation \eqref{ann:en_sum} is a quickly
converging function so it should be easy to evaluate. However it can
be difficult to evaluate the $\eta$ eigenvalues, which become
functions of the wavenumber $\kappa$ and a natural number $m$. We can
get around this problem by using \eqref{ann:en_int}. We cannot
integrate along the real line because of the poles introduced when we
used the argument principle, and we cannot distort the contour
to one running  along the
imaginary axis because the integral then becomes divergent. So a
simple choice is then to let the $\eta$ integration run along the angles of
$\pi/4$ and $-\pi/4$. Writing $\tr
g_\eta^{(1)}v_1g_\eta^{(2)}v_2=A(\eta)$ we have
{\small \begin{multline}\label{num:int} \frac{E}{L_z}=-\frac{1}{4\pi^2}
    \int_0^\infty \kappa d \kappa \int_0^\infty \dif\nu \\ \times\Bigg\{
    \frac{\Re R_{\sqrt{i}\nu}\partial_\nu\Re R_{\sqrt{i}\nu}+ \Im
      R_{\sqrt{i}\nu}\partial_\nu\Im R_{\sqrt{i}\nu}}{
      \left|R_{\sqrt{i}\nu}\right|^2} \arctan\left(\frac{\Im
      A(\sqrt{i}\nu)}{1-\Re A(\sqrt{i}\nu)}\right)\\ 
\mbox{}-\frac{\Re
      R_{\sqrt{i}\nu}\partial_\nu\Im R_{\sqrt{i}\nu}- \Im
      R_{\sqrt{i}\nu}\partial_\nu\Re R_{\sqrt{i}\nu}}{2
      \left|R_{\sqrt{i}\nu}\right|^2} \\
\times\ln\left(1-2 \Re
    A(\sqrt{i}\nu)+\left|A(\sqrt{i}\nu)\right|^2\right) \Bigg\}.
\end{multline} }
Here we have used the property that $R_{\eta^*}=R_\eta^*$, and
$A(\eta^*)=A^*(\eta)$. The value of $R_{\sqrt{i}\nu}(b,\kappa)$ is
obtained as the numerical solution of the differential
equation. Using this technique we can obtain a numerical energy in
about 1 cpu-second. The results of this calculation are found in
figure \ref{ann:pi_4}.

Again we would like to compare to known results, so figure
\ref{ann:comparison} is a graph of the ratio of the energies of an
annular piston, and a rectangular piston of similar dimension. The
rectangular piston is constructed so it has the same finite width
$b-a$ as the annular piston, and the separation distance is the mean
distance between the annular plates,
\begin{equation}\label{num:mean_dist}
d=\frac{b+a}{2}2\sin\left(\frac{\alpha}{2}\right).
\end{equation}
The results make a certain amount of physical sense. The energy of the
annular piston is greater than that of the rectangular piston for
small separation because the inner edge of the annular piston is
closer, and will contribute more to the energy. However as the annular
piston gets further away, the other side of the piston will start to
contribute and lower the overall energy. In addition we see that the
energy for a small  piston is much closer to 
that of the rectangular piston for small
separations than for a larger piston,
$E_{\text{ann}}/E_{\text{rect}}\approx 1.004$ for $b/a=1.1$
vs. $E_{\text{ann}}/E_{\text{rect}}\approx 1.23$ for $b/a=2$. In both
cases the value approached in the plateau in figure
\ref{ann:comparison} is very close to the ratio of the energies of a
flat plate to that of a tilted plate predicted by using 
the proximity force approximation.

\section*{Acknowledgments}
This material is based upon work supported by the
National Science Foundation under Grants Nos.~PHY-0554926 (OU) and
PHY-0757791 (BU) and by the
US Department of Energy under Grants Nos.~DE-FG02-04ER41305 and DE-FG02-04ER-%
46140 (both OU).
We thank Simen Ellingsen, Iver Brevik, Prachi Parashar,
Nima Pourtolami, and Elom Abalo for collaboration.
Part of the work was done while KK enjoyed the hospitality and partial
support of the
Department of Physics and Astronomy of the University of Oklahoma.
Thanks go in particular to Kimball Milton and his group who
made this very pleasant and exciting visit possible.

\bibliography{wagner_qfext09.bib}

\end{document}